# Benchmarking a Grid Foundation Model for AC Optimal Power Flow: Feasibility and Warm-Start Results on Synthetic Texas and South Carolina Grids

Tam Kemabonta Brighton Ombuki Vladimir Abdelnour

*Abstract*—**Learning-based surrogates for alternating-current optimal power flow (AC-OPF) are often judged by the gap between their objective value and that of a nonlinear solver, which does not show that a prediction satisfies power balance and network limits. This paper tests whether predictions from the pretrained foundation model GridSFM-Open 1.1.0, recently released by Microsoft, are feasible AC-OPF operating points and whether they are useful initial values for a nonlinear solver. The model was evaluated zero-shot and after fine-tuning on 10, 100, and 1,000 labeled scenarios (FT10, FT100, and FT1000) on 500 scenarios for each of two synthetic grids, ACTIVSg2000 (Texas) and ACTIVSg500 (South Carolina), with Ipopt, MATPOWER MIPS, and pandapower PIPS as solver references. The lowest Texas median cost gap, 0.44% for FT100, coincided with thermal violations in 24.1% of cases and a median maximum bus power mismatch of 89 MW and 197 Mvar. Texas zero-shot predictions had a median maximum mismatch of 648 MW and 1,641 Mvar, and every South Carolina case violated a thermal limit for every model variant. Seeding Ipopt with FT1000 predictions gave geometric-mean solve-time speedups of 1.55× on South Carolina and 1.12× on Texas ($p = 0.06$, not significant at the 5% level), while zero-shot seeds slowed South Carolina solves (0.26×). A small objective gap therefore did not indicate feasibility: although median zero-shot inference took only 1.32 s on Texas and 0.157 s on South Carolina on a shared host, every prediction still required a nonlinear solver, and fine-tuned predictions gave modest, grid-dependent warm-start speedups while zero-shot predictions slowed convergence on one grid. GridSFM may therefore be useful for initializing a nonlinear solver, but obtaining feasible AC-OPF solutions still required the solver.**

## I. INTRODUCTION

Alternating-current optimal power flow (AC-OPF) is the nonlinear program that sets cost-minimizing generator dispatch subject to power-balance, voltage, thermal, and angle limits. Open nonlinear solvers such as Ipopt through PowerModels [1], [2], the MATPOWER Interior Point Solver (MIPS) [3], and the Python Interior Point Solver (PIPS) used by pandapower [4] remain the reference for research and education. Learning-based surrogates have shortened wall-clock time on fixed topologies [5], [6]; GridSFM extends that line with a single pretrained foundation model that runs zero-shot across topologies and adapts with few-shot fine-tuning [7], [8], [9].

An AC-OPF solution must satisfy the power-balance equations and the network limits, so this paper measures feasibility alongside the objective gap. The question addressed here is whether GridSFM predictions are feasible operating points and, where they are not, whether they still help a nonlinear solver.

This paper reports a central processing unit (CPU) benchmark of GridSFM-Open 1.1.0 against the open solvers on two public synthetic grids: ACTIVSg2000 on the footprint of Texas and ACTIVSg500 on the footprint of South Carolina [10], [11], [12]. The contributions are:

- A feasibility assessment of GridSFM predictions, reported as power-balance mismatch and thermal, voltage, angle, and generator-limit violations next to the objective gap, for zero-shot and fine-tuned models on 500 scenarios per grid.
- A warm-start test in which GridSFM predictions initialize Ipopt, compared with a cold start under identical solver options.
- Timing, solver-agreement, and infeasibility-detection results, reported together with the run conditions that qualify them.

## II. RELATED WORK

Open AC-OPF stacks expose shared nonlinear formulations through PowerModels [1], MATPOWER [3], and pandapower [4], typically driven by Ipopt [2] or interior-point or primal-dual solvers such as MIPS and PIPS. Neural OPF approaches include DeepOPF for security-constrained DC-OPF [5] and DeepOPF-V for AC-OPF [6]. GridSFM [7], [8] is an approximately 15 million-parameter graph foundation model pretrained across dozens of topologies and released with open checkpoints and code [9]; its training corpus draws on public cases and related OPF datasets [13]. The model card reports about 540,000 pretraining scenarios on 54 base topologies, namely 24 pglib-OPF transmission cases of 500 to 4,661 buses and 30 multi-state planning-region grids [14]. The card does not list the topologies by name, so whether ACTIVSg2000 or ACTIVSg500 was part of the pretraining data is not established.

## III. EXPERIMENT DESIGN

### A. Grids, tools, and formulation

The study evaluated ACTIVSg2000 (2,000 buses, Texas footprint) and ACTIVSg500 (500 buses, South Carolina footprint). GridSFM-Open 1.1.0 was loaded from the microsoft/gridsfm tree at commit 1ca775f [9]. The solver references were PowerModels 0.21.7 with Ipopt.jl 1.16.0 under Julia 1.11.7, MATPOWER 8.1.1-dev under Octave (MIPS), and pandapower 3.5.6 (PIPS). A direct-current OPF (DC-OPF) reference was solved in PowerModels. The harmonized AC-OPF used voltage limits of 0.90 to 1.10 per unit (p.u.), thermal rate A, and angle-difference limits of ±60°. All runs were CPU-only on one virtual machine with 8 cores of an Intel Xeon processor, 15 GB of RAM, and no GPU.

Stopping criteria were solver specific. Ipopt used an optimality tolerance and a constraint-violation tolerance of $10^{-8}$ each and a limit of 3,000 iterations. MIPS used feasibility, gradient, complementarity, and cost tolerances of $10^{-8}$ and the default limit of 150 iterations. PIPS used the pandapower 3.5.6 defaults: an OPF violation tolerance of $5 \times 10^{-6}$, gradient, complementarity, and cost tolerances of $10^{-6}$, and 150 iterations. No solver had a time limit in the benchmark runs. Each solver applied its own convergence criteria, as listed above. A run counted as converged when the solver reported a locally solved, optimal, almost locally solved, or success status.

### B. Scenarios, fine-tuning, and validation

Each grid had 500 evaluation scenarios spanning seven perturbation families which includes, global load scaling, per-bus load perturbation, single-branch outage, single-generator outage, branch-rating derating, renewable availability, and combined stress.

Fine-tuning scenarios were generated separately, with different master random seeds, and each was labeled with an Ipopt solution. In manifest order, the first 1,000 scenarios with a converged label formed the training pool, and the following converged scenarios (93 for Texas and 85 for South Carolina) formed a post-pool validation list. Nested family-stratified subsets of 10 and 100 scenarios were drawn from the pool; the models fine-tuned on them are denoted FT10 and FT100, the model fine-tuned on all 1,000 pool scenarios is denoted FT1000, and the unmodified pretrained model is denoted zero-shot. Fine-tuning ran on the CPU for up to 10 epochs (AdamW, learning rate $10^{-4}$, weight decay $10^{-4}$, batch size 1, gradient-norm clipping at 5) and kept the epoch with the lowest validation median cost gap.

The validation protocol differed with training size. For FT10 and FT100, validation used 100 pool scenarios outside the training subset. For FT1000, which used the whole pool, validation used the post-pool list. Both protocols scored the cost gap on the first 40 validation scenarios. On Texas, those 40 post-pool scenarios were 39 renewable-availability scenarios and one combined-stress scenario. The Texas FT1000 run stopped after four epochs and used the epoch-1 checkpoint (validation median cost gap 0.51%). South Carolina FT1000 used the 85-scenario post-pool list and selected epoch 4 (validation median cost gap 1.46%).

Evaluation and fine-tuning scenarios came from separate generator runs with no shared random seeds; FT10 and FT100 validation was held out from the fine-tuning pool, and FT1000 validation used the post-pool remainder of the same fine-tuning generation. Single-branch and single-generator outage scenarios have a finite set of possible outage choices, however, and 24 of the 500 Texas evaluation scenarios (8 branch and 16 generator outages) and 91 of the 500 South Carolina evaluation scenarios (42 branch and 49 generator outages) are identical to a scenario in the fine-tuning manifest. Of these, 24 and 85, respectively, lie in the 1,000-scenario training pool. Excluding the identical scenarios changed each median cost gap in Table I by no more than 0.31 percentage points.

### C. Cost-gap and feasibility metrics

For each scenario on which Ipopt converged and a prediction was returned, the cost gap was the unsigned relative difference

$$g = \frac{|f_{\text{pred}} - f_{\text{Ipopt}}|}{|f_{\text{Ipopt}}|} \tag{1}$$

where $f_{\text{Ipopt}}$ is the converged Ipopt objective and $f_{\text{pred}}$ is the quadratic generator cost evaluated at the dispatch returned by the method. The common subset was the 498 Texas and 488 South Carolina scenarios on which Ipopt converged; for MIPS and PIPS it was the subset on which that solver also converged. Medians over scenarios are reported.

Every method was scored by the same evaluator from its own voltages, angles, and dispatch. The power-balance mismatch at bus $k$ was

$$\Delta S_k = S_{g,k} - S_{d,k} - S_{\text{base}} V_k \left(\textstyle\sum_m Y_{km} V_m\right)^* \tag{2}$$

with generation $S_{g,k}$ and load $S_{d,k}$ in MVA, voltages $V$ and admittances $Y$ in p.u., and $S_{\text{base}} = 100$ MVA. The real and imaginary parts of $\Delta S_k$ are reported in MW and Mvar as the maximum absolute value over buses, and the median of that maximum over scenarios is given. No pass threshold was applied to the mismatch. The loading ratio of branch $b$ was

$$\ell_b = \frac{\max(|S_{f,b}|, |S_{t,b}|)}{S_b^{\max}} \tag{3}$$

and a thermal violation was recorded when $\ell_b > 1.001$ for the rate-A limit $S_b^{\max}$. A voltage violation was a magnitude beyond its limit by more than $10^{-4}$ p.u., an angle violation was a branch angle difference above 60.01°, and a generator violation was active or reactive output beyond its limit by more than 0.1 MW or 0.1 Mvar. The mismatch was not scored for PIPS because its saved dispatch was incomplete.

### D. Warm-start test and timing conditions

On a pinned subset of 50 evaluation scenarios per grid, Ipopt in PowerModels was seeded with initial voltage magnitudes, voltage angles, active power, and reactive power (vm_start, va_start, pg_start, and qg_start). The seeds were GridSFM zero-shot predictions, GridSFM FT1000 predictions, ground-truth (GT) seeds, defined as the converged cold-start Ipopt solution of the same scenario, and DC-OPF solutions. Each warm run was paired with a cold run under identical Ipopt options and a 120 s time limit. The speedup of a scenario was $t_{\text{cold}}/t_{\text{warm}}$. The geometric mean over jointly converged pairs is reported, with a two-sided Wilcoxon signed-rank test on the paired times. Two objectives matched when their relative difference was at most $10^{-4}$. A control run enabled Ipopt warm-start options but applied no start values, because PowerModels ignored the keys that were written; it isolates the effect of those options.

CPU-pinned timing used one-thread settings and separate CPU affinities: PowerModels and GridSFM ran sequentially on core 4 while MATPOWER and pandapower ran concurrently on cores 5 and 6. The host was shared, and Texas FT1000 training continued on other cores during the measurements. GridSFM inference excluded model loading and followed a discarded warm-up prediction. Reported times covered solver or inference calls rather than an end-to-end workflow.

## IV. RESULTS

### *A. Convergence and solver agreement*

Of the 500 evaluation scenarios, Ipopt and MIPS each converged on 498 Texas and 488 South Carolina scenarios, PIPS on 495 and 494, and DC-OPF on 498 and 500 (Table II). GridSFM produced a prediction for all 500 scenarios on both grids, and the feasibility of those predictions is reported in Section IV-B. Ipopt and MIPS agreed on the objective to a median relative difference of $8.9 \times 10^{-10}$ on Texas and $1.3 \times 10^{-8}$ on South Carolina. PIPS differed from Ipopt by $2.2 \times 10^{-5}$ on Texas and $9.0 \times 10^{-3}$ on South Carolina, where the difference exceeded 0.1% in 99% of the jointly converged cases and PIPS solutions exceeded branch ratings at the evaluator thresholds in 98.8% of cases (Table I). Ipopt was therefore the single reference for cost gaps. DC-OPF had median gaps of 2.24% on Texas and 2.42% on South Carolina.

### *B. Objective gap and feasibility of predictions*

Table I and Fig. 1 report the objective gap together with the feasibility of each prediction. The zero-shot median gap was 5.36% on Texas and 46.0% on South Carolina. After fine-tuning, median gaps were between 0.44% and 1.36% on Texas and between 1.08% and 8.56% on South Carolina. GridSFM predictions had no voltage, angle, or generator-limit violations at the evaluator thresholds, but power balance was not satisfied and thermal limits were exceeded.

The median maximum bus mismatch of Texas zero-shot predictions was 648 MW and 1,641 Mvar, and that of Texas FT1000 predictions was 56 MW and 182 Mvar, whereas Ipopt and MIPS solutions had mismatches below 0.001 MW and 0.001 Mvar. The lowest Texas median gap, 0.44% for FT100, coincided with a median maximum mismatch of 89 MW and 197 Mvar and with thermal violations in 24.1% of cases; Texas FT1000 predictions had thermal violations in 17.9% of cases. On South Carolina, every case of every GridSFM variant had a thermal violation, with a median of 149 overloaded branches for zero-shot predictions and 32 for FT1000 predictions. Violation shares used the scenarios on which Ipopt converged.

TABLE I. MEDIAN COST GAP AND FEASIBILITY OF PREDICTIONS (EVALUATION SETS)

| Method | Cost gap (%) | Max. ΔP (MW) | Max. ΔQ (Mvar) | Thermal violations (% of cases) |
|---|---|---|---|---|
| **Texas (ACTIVSg2000)** | | | | |
| Ipopt | reference | <0.001 | <0.001 | 0.0 |
| MIPS | <0.001 | <0.001 | <0.001 | 0.0 |
| PIPS | 0.0022 | n/a | n/a | 0.6 |
| GridSFM zero-shot | 5.36 | 648 | 1,641 | 100 |
| GridSFM FT10 | 0.866 | 164 | 534 | 99.8 |
| GridSFM FT100 | 0.442 | 89.0 | 197 | 24.1 |
| GridSFM FT1000 | 1.357 | 56.5 | 182 | 17.9 |
| **South Carolina (ACTIVSg500)** | | | | |
| Ipopt | reference | <0.001 | <0.001 | 0.0 |
| MIPS | <0.001 | <0.001 | <0.001 | 0.0 |
| PIPS | 0.90 | n/a | n/a | 98.8 |
| GridSFM zero-shot | 46.0 | 325 | 872 | 100 |
| GridSFM FT10 | 4.51 | 166 | 433 | 100 |
| GridSFM FT100 | 8.56 | 59.2 | 59.6 | 100 |
| GridSFM FT1000 | 1.084 | 56.0 | 28.3 | 100 |

ΔP and ΔQ: median over scenarios of the maximum absolute bus mismatch. n/a: not scored. FT10 and FT100 used different validation sets and protocol than FT1000 (Section IV-C).

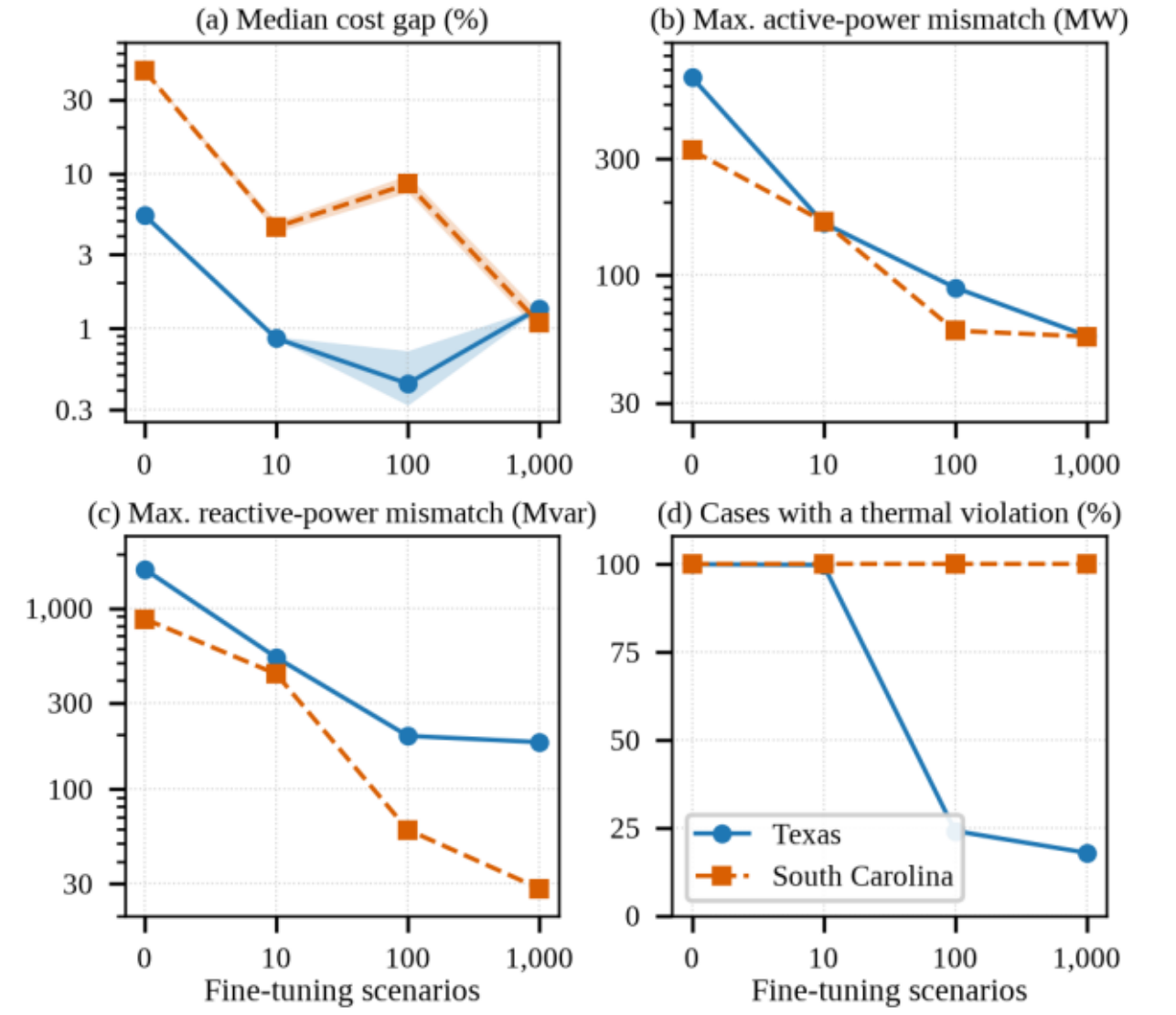

Figure 1. GridSFM predictions by number of fine-tuning scenarios (0 denotes zero-shot); each point uses the validation protocol given in Section IV-C: (a) median cost gap versus Ipopt, with shaded bands showing 95% bootstrap confidence intervals of the median (10,000 resamples); (b) median maximum bus active-power mismatch; (c) median maximum bus reactive-power mismatch; (d) share of scenarios with at least one thermal violation.

### C. Interpretation of the fine-tuning trend

Cost gaps varied non-monotonically with the number of fine-tuning scenarios (Fig. 1(a)), reflecting the different validation protocols in Section IV-C. First, the model-selection protocol changed with size. On Texas, the zero-shot median gap on the FT1000 validation list was 0.31%, against 5.3% on the FT10 and FT100 validation sets, so the lists differ in difficulty. Second, alternative choices recorded for South Carolina moved the medians: FT10 selected on the 85-scenario post-pool validation list gave 10.45% instead of 4.51%, FT100 trained on the first scenarios in manifest order, all load-scaling scenarios, gave 2.19% instead of 8.56%, and the final-epoch FT1000 checkpoint gave 1.95% instead of 1.08%. The mismatch also decreased with training size (Fig. 1(b) and 1(c)) while the thermal-violation share fell only on Texas (Fig. 1(d)).

### D. Warm start of Ipopt

Warm-starting Ipopt was the setting in which GridSFM predictions were most useful. With FT1000 predictions as the starting point, Ipopt solved the South Carolina subset with a geometric-mean speedup of 1.55× ($p < 10^{-10}$, 90% of scenarios faster), and median iterations fell from 45 to 30.5. The median solve time fell from 0.583 s to 0.364 s. Zero-shot seeds, in contrast, slowed Ipopt on South Carolina (0.26×, with 4% of scenarios faster). GT and DC-OPF seeds gave 1.49× and 1.48× on South Carolina (Fig. 2).

On Texas, the FT1000 geometric-mean speedup was 1.12×, and the median solve time fell from 6.63 s to 5.66 s. Of the 50 Texas FT1000 runs, 49 converged in both the warm and the cold run; one warm run exceeded the 120 s limit (recorded solve time 159.8 s) and was excluded from the speedup and from the test. The Wilcoxon test on the 49 pairs gave $p = 0.06$, which does not meet the 5% significance level, and counting the timed-out pair at its recorded warm time lowers the geometric mean to 1.05× ($p = 0.11$). The other Texas seeds gave 1.22× (zero-shot) and 1.17× (GT and DC-OPF) with $p < 10^{-5}$, so FT1000 seeds gave the smallest Texas speedup of the four seed types. A control run with Ipopt warm-start options enabled but no start values applied gave 0.57×, which confirms that the speedups come from the supplied start values. Warm and cold runs reached matching objectives on every jointly converged pair. The speedups compare Ipopt solve times only and exclude the time needed to produce the seed.

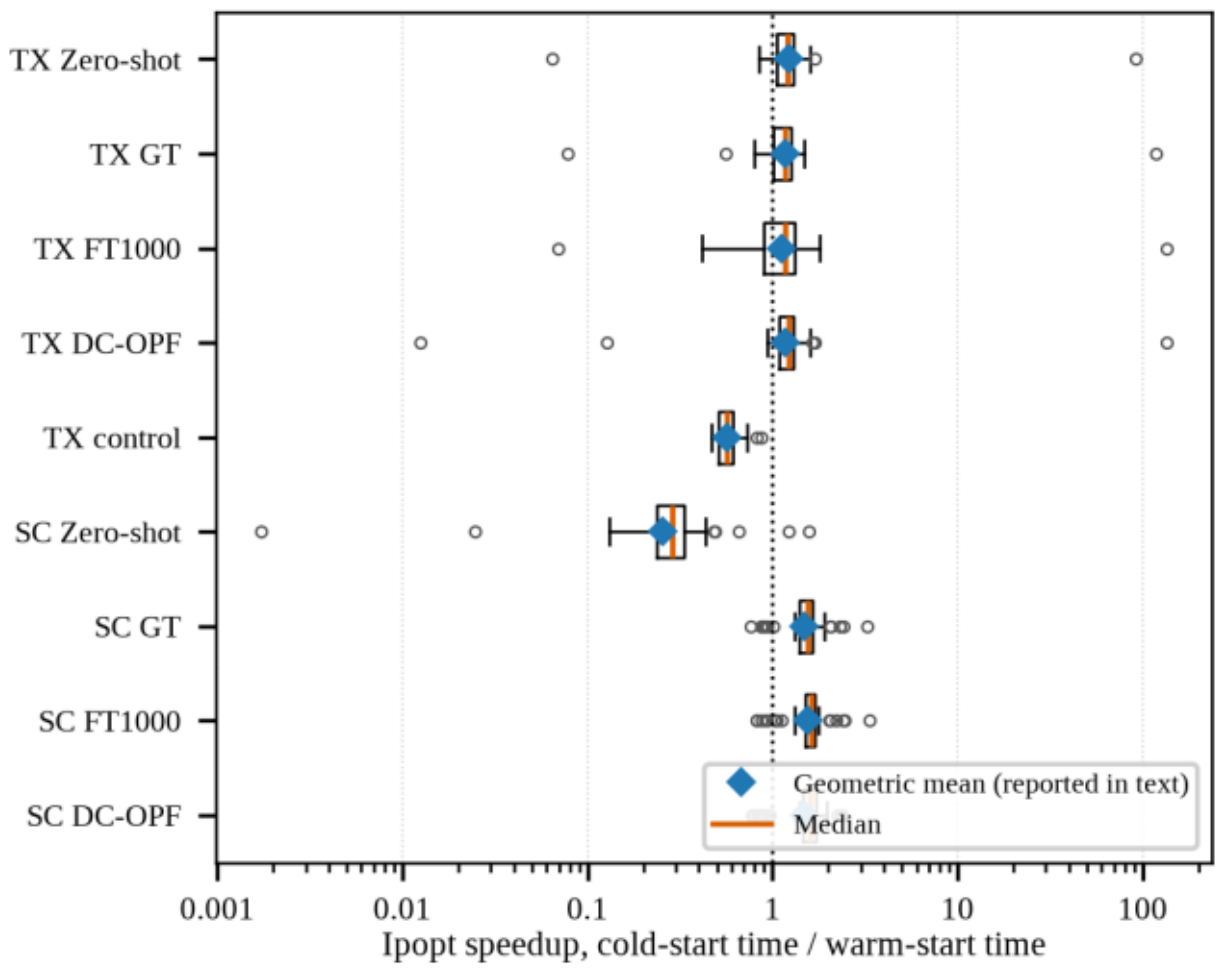

Figure 2. Distribution of Ipopt warm-start speedups (cold-start time divided by warm-start time) over jointly converged scenarios of the pinned 50-scenario subsets. Boxes show quartiles and median, whiskers extend to 1.5 times the interquartile range, circles are outliers, and diamonds mark the geometric means given in the text. GT: ground-truth seed. Control: Ipopt warm-start options with no start values applied (Texas only).

### E. Timing on the shared host

Table II lists the pinned-subset medians. GridSFM zero-shot inference took 1.32 s on Texas and 0.157 s on South Carolina, against 6.25 s and 0.558 s for Ipopt, 3.39 s and 0.674 s for MIPS, and 21.4 s and 1.24 s for PIPS (Fig. 3). These times were measured under the conditions in Section III: the host was shared, FT1000 training ran on other cores, the solvers ran in Julia, Octave, and Python implementations, and the GridSFM times cover inference only, before any step to make the predictions feasible.

TABLE II. CONVERGENCE COUNTS (OF 500 SCENARIOS) AND PINNED-SUBSET MEDIAN TIME

| Method | Texas converged | Texas time (s) | South Carolina converged | South Carolina time (s) |
|---|---|---|---|---|
| Ipopt | 498 | 6.25 | 488 | 0.558 |
| MIPS | 498 | 3.39 | 488 | 0.674 |
| PIPS | 495 | 21.4 | 494 | 1.24 |
| DC-OPF | 498 | n/a | 500 | n/a |
| GridSFM zero-shot* | 500 | 1.32 | 500 | 0.157 |

*GridSFM: predictions returned.*

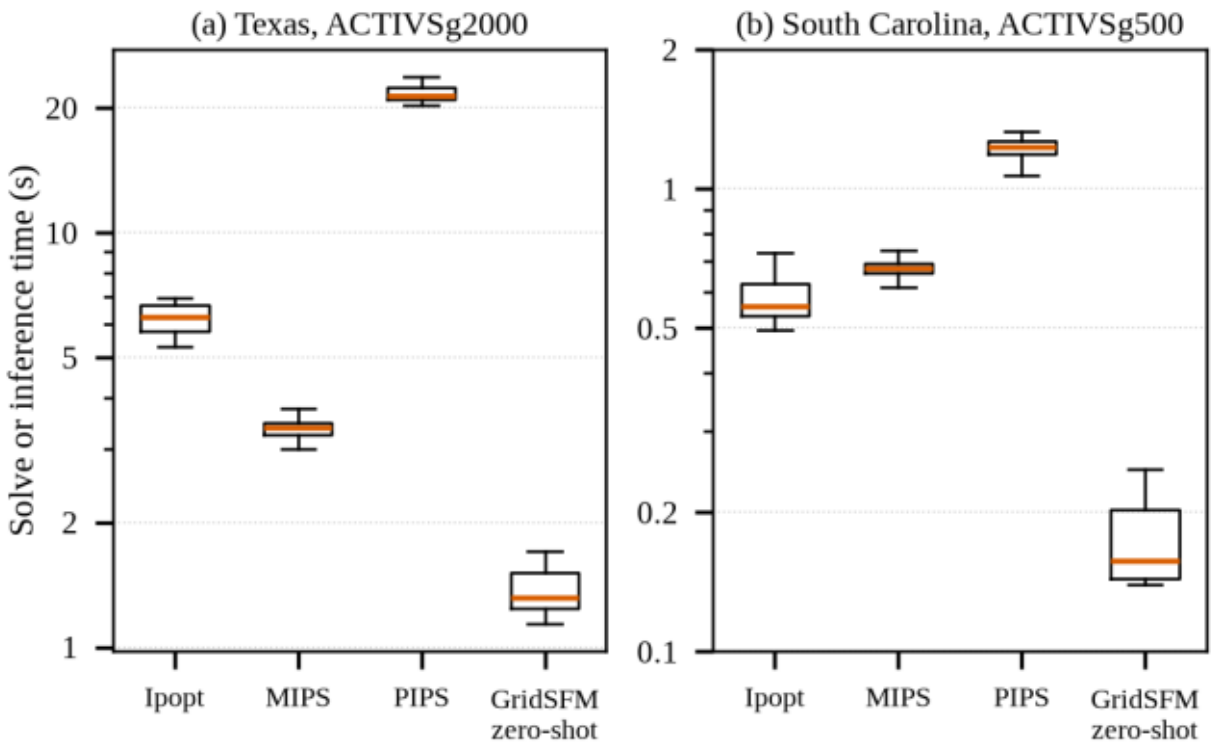

Figure 3. Pinned-subset solve or inference time on (a) Texas and (b) South Carolina (50 scenarios each; 49 successful records for Texas MIPS and PIPS). Boxes show quartiles and median, and whiskers extend to 1.5 times the interquartile range. GridSFM times exclude model loading.

### F. *Infeasibility labels*

Ipopt did not converge on 2 Texas and 12 South Carolina scenarios, and MIPS did not converge on the same scenarios. These nonconvergence were used as positive labels for infeasibility. The labels mark local-solver nonconvergence and are therefore proxies for infeasibility. The area under the receiver operating characteristic curve (ROC-AUC) of the GridSFM feasibility output on Texas rose from 0.42 (zero-shot) to 0.95 (FT100) and was 0.88 for FT1000, but it rests on only 2 positive scenarios, and no confidence interval was computed. The South Carolina values were 0.50, 0.59, 0.65, and 0.78 for zero-shot, FT10, FT100, and FT1000 with 12 positives. At the threshold of 0.5 on the feasibility output, no Texas variant flagged any scenario, and South Carolina FT100 and FT1000 flagged none. South Carolina zero-shot flagged all 500 scenarios, and South Carolina FT10 flagged 10 of the 12 positives and 311 of the 488 others.

## V. DISCUSSION AND CONCLUSION

GridSFM-Open returned predictions with median zero-shot inference times of 1.32 s on Texas and 0.157 s on South Carolina on a shared host (Section IV-E) and, after fine-tuning, with median objective gaps of 0.44% to 1.36% on Texas. The predictions violated AC-OPF constraints: the median maximum bus mismatch ranged from 56 to 648 MW and from 28 to 1,641 Mvar across variants and grids, thermal limits were exceeded in 17.9% to 100% of Texas cases and in all South Carolina cases, and a nonlinear solver remained necessary to obtain a feasible solution. When a prediction served as the initial point, Ipopt solved faster on average, by 1.55× on South Carolina and 1.12× on Texas for FT1000, although the Texas result was not significant at the 5% level. Zero-shot South Carolina predictions slowed Ipopt (0.26×). On South Carolina, GT and DC-OPF seeds gave 1.49× and 1.48× against 1.55× for FT1000 seeds, and on Texas the FT1000 seeds (1.12×) gave less than the GT and DC-OPF seeds (1.17×) and the zero-shot seeds (1.22×). FT1000 seeds gave speedups similar to DC-OPF seeds on South Carolina (1.55× and 1.48×) and smaller ones on Texas (1.12× and 1.17×).

The results suggest that GridSFM may be useful for initializing nonlinear solvers. Using it as a direct source of AC-OPF solutions would require closing the feasibility gap, for example with a feasibility-restoration step, which remains future work. The evidence is limited to two synthetic grids, 500 evaluation scenarios each, 50-scenario warm-start subsets, one shared host, identical outage scenarios in the evaluation and fine-tuning sets, an undetermined overlap between the evaluation grids and the pretraining data, and an infeasibility-label analysis (Section IV-F) that rests on 2 Texas positives.

## References

[1] C. Coffrin, R. Bent, K. Sundar, Y. Ng, and M. Lubin, "PowerModels.jl: An open-source framework for exploring power flow formulations," in *Proc. Power Syst. Comput. Conf. (PSCC)*, Jun. 2018, pp. 1–8, doi: 10.23919/PSCC.2018.8442948.

[2] A. Wächter and L. T. Biegler, "On the implementation of an interior-point filter line-search algorithm for large-scale nonlinear programming," *Math. Program.*, vol. 106, no. 1, pp. 25–57, Mar. 2006, doi: 10.1007/s10107-004-0559-y.

[3] R. D. Zimmerman, C. E. Murillo-Sánchez, and R. J. Thomas, "MATPOWER: Steady-state operations, planning, and analysis tools for power systems research and education," *IEEE Trans. Power Syst.*, vol. 26, no. 1, pp. 12–19, Feb. 2011, doi: 10.1109/TPWRS.2010.2051168.

[4] L. Thurner *et al.*, "pandapower—An open-source Python tool for convenient modeling, analysis, and optimization of electric power systems," *IEEE Trans. Power Syst.*, vol. 33, no. 6, pp. 6510–6521, Nov. 2018, doi: 10.1109/TPWRS.2018.2829021.

[5] X. Pan, T. Zhao, M. Chen, and S. Zhang, "DeepOPF: A deep neural network approach for security-constrained DC optimal power flow," *IEEE Trans. Power Syst.*, vol. 36, no. 3, pp. 1725–1735, May 2021, doi: 10.1109/TPWRS.2020.3026379.

[6] W. Huang, X. Pan, M. Chen, and S. H. Low, "DeepOPF-V: Solving AC-OPF problems efficiently," *IEEE Trans. Power Syst.*, vol. 37, no. 1, pp. 800–803, Jan. 2022, doi: 10.1109/TPWRS.2021.3114092.

[7] W. Yang, A. Britto Mattos Lima, T. V. Spina, S. Fowers, B. Zhang, and C. White, "GridSFM: A foundation model for AC optimal power flow," Microsoft Research, technical white paper, May 2026. [Online].

[8] L. Bhan, W. Yang, M. Capetz, and B. Zhang, "GridSFM: A foundation model for solving AC optimal power flow," 2026, *arXiv:2609.30173*.

[9] Microsoft, "GridSFM: Open-source AC optimal power flow foundation model," GitHub repository, commit 1ca775f, gridsfm 1.1.0, 2026. [Online]. Available: https://github.com/microsoft/GridSFM

[10] A. B. Birchfield, T. Xu, K. M. Gegner, K. S. Shetye, and T. J. Overbye, "Grid structural characteristics as validation criteria for synthetic networks," *IEEE Trans. Power Syst.*, vol. 32, no. 4, pp. 3258–3265, Jul. 2017, doi: 10.1109/TPWRS.2016.2616385.

[11] Texas A&M University, "ACTIVSg2000: 2000-bus synthetic grid on footprint of Texas," Electric Grid Test Case Repository. [Online]. https://electricgrids.engr.tamu.edu/electric-grid-test-cases/activsg2000/

[12] Texas A&M University, "ACTIVSg500: 500-bus synthetic grid on footprint of South Carolina," Electric Grid Test Case Repository. [Online]. Available: https://electricgrids.engr.tamu.edu/electric-grid-test-cases/activsg500/

[13] S. Lovett *et al.*, "OPFData: Large-scale datasets for AC optimal power flow with topological perturbations," 2024, *arXiv:2406.07234*.

[14] Microsoft, "Grid Small Foundation Model (GridSFM)," Hugging Face model card, 2026. [Online]. Available: https://huggingface.co/microsoft/GridSFM_Open